# Introduction to Distributed Systems

SABU M. THAMPI
L.B.S INSTITUTE OF TECHNOLOGY FOR WOMEN
TRIVANDRUM, KERALA, INDIA-695012
smtlbs@in.com, smtlbs@gmail.com

---

**Abstract:** Computing has passed through many transformations since the birth of the first computing machines. Developments in technology have resulted in the availability of fast and inexpensive processors, and progresses in communication technology have resulted in the availability of lucrative and highly proficient computer networks. Among these, the centralized networks have one component that is shared by users all the time. All resources are accessible, but there is a single point of control as well as a single point of failure. The integration of computer and networking technologies gave birth to new paradigm of computing called *distributed computing* in the late 1970s. Distributed computing has changed the face of computing and offered quick and precise solutions for a variety of complex problems for different fields. Nowadays, we are fully engrossed by the information age, and expending more time communicating and gathering information through the Internet. The Internet keeps on progressing along more than a few magnitudes, abiding end systems increasingly to communicate in more and more different ways. Over the years, several methods have evolved to enable these developments, ranging from simplistic data sharing to advanced systems supporting a multitude of services. This article provides an overview of distributed computing systems. The definition, architecture, characteristics of distributed systems and the various distributed computing fallacies are discussed in the beginning. Finally, discusses client/server computing, World Wide Web and types of distributed systems.

## What is a Distributed System?

The scale of networked workstations and plunge of the centralized mainframe has been the most theatrical change in the last two decades of information technology. This shift has placed added processing power in the hands of the end-user and distributed hardware resources. When computers work together over a network, the network may use the power from all the networked computers to perform complex tasks. Computation in networks of processing nodes can be classified into *centralized or distributed computations*. A centralized solution relies on one node being designated as the computer node that processes the entire application locally and the central system is shared by all the users all the time. Hence there is single point of control and single point of failure.

The motivation behind the growth of decentralised computing is the availability of the low-priced, high performance computers and network tools. When a handful of powerful computers are linked together and communicate with each other, the overall computing power available can be amazingly vast. Such a system can have a higher performance share than a single supercomputer. Distributed computing – a decentralisation approach to computing is a potentially very powerful approach for accessing large amounts of computational power. The objective of such systems is to minimize communication and computation cost. In distributed systems, the processing steps of the application are divided among the participating nodes. The basic step in all distributed computing architectures is the notion of communication between computers.

*Distributed system is an application that executes a collection of protocols to coordinate the actions of multiple processes on a communication network, such that all components cooperate together to perform a single or small set of related tasks.* The collaborating computers can access remote resources as well as local resources in the distributed system via the communication network. The existence of multiple autonomous computers is transparent to the user in a distributed system. The user is not aware that the jobs are executed by multiple computers subsist in remote locations. This means that like centralised systems no single computer in the system carries the entire load on system resources that running a computer program usually required.

## Distributed System Architecture

Distributed systems are built up on top of existing networking and operating systems software. A distributed system comprises a collection of autonomous computers, linked through a computer network and distribution middleware. To become autonomous there exist a clear master/slave association between two computers in the network. The middleware enables computers to coordinate their activities and to share the resources of the system, so that users perceive the system as a single, integrated computing facility. Thus, middleware is the bridge that connects distributed applications across dissimilar physical locations, with dissimilar hardware platforms, network technologies, operating systems, and programming languages. The middleware software is being developed following agreed standards and protocols. It provides standard services such as naming, persistence, concurrency control to ensures that accurate results for concurrent processes are produced and obtains the results as fast as possible, event distribution, authorization to specify access rights to resources, security etc. The middleware service extends over multiple machines. Figure 1 shows a simple architecture of a distributed system [GEOR01, ANDR02].

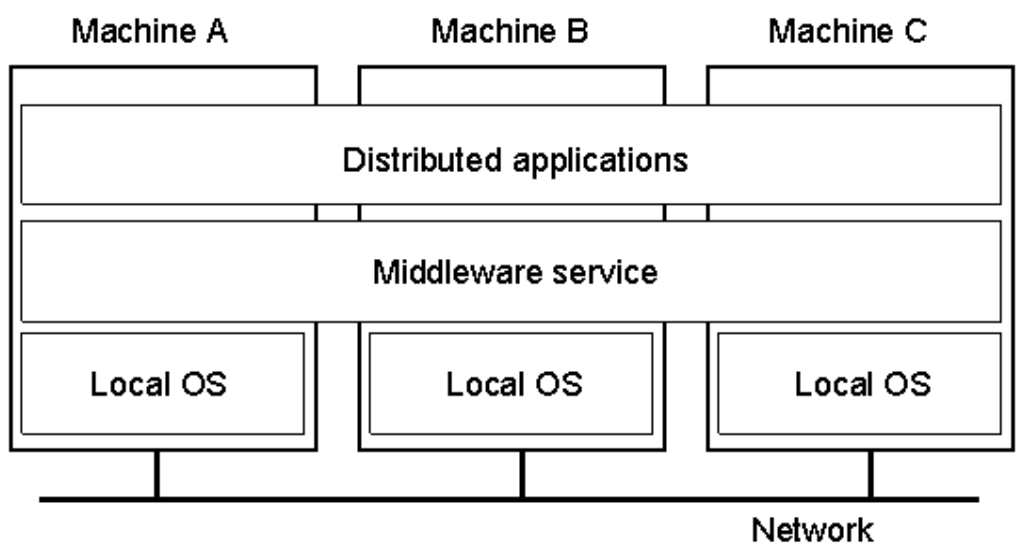

Figure 1. A Distributed System

The distributed system can be viewed as defined by the physical components or as defined from user or computation point of view. The first is known as the physical view and the second as the logical view. Physically a distributed system consists of a set of nodes (computers) linked together by a communication network. The nodes in the network are loosely coupled and do not share their memory. The nodes in the system communicate by passing messages over the communication network. Communication protocols are used for sending messages from one node to another. The logical model is the view that an application has of the system. It contains a set of concurrent processes and communication channels between them. The core network is treated as fully connected. Processes communicate by sending messages to each other. A system is synchronous if during a proper execution, it all the time performs the intended operation in a known fixed time, otherwise it is asynchronous. In synchronous system the failure can be noticed by a lack of response from the system. Therefore, timeout based techniques are used for failure discovery.

A distributed system can be constructed by means of fully connected networks or partially connected networks [HUAN05, MIN04, NELS01]. A fully connected network (figure 2) is a network in which each of the nodes is connected to each other. The problem with such a system is that adding new nodes to the system results in the increase of number of nodes connected to the node. Due to this the number of file descriptors and complexity for each node to implement the connections are increased heavily. Thus, the scalability (capability of a system to continue to function well when the system is changed in size or volume) of such systems is limited by each node's capacity to open file descriptors and the ability to handle the new connections. The communication cost - the message delay of sending a message from the source to the destination- is low because a message sent from one computer to another one only goes through one link. Fully connected systems are reliable because when a few computers or links fail, the rest of the computers can still communicate with others.

In a partially connected network, direct links exist between some, but not all, pairs of computers. A few of the partially connected network models are star structured networks, multi-access bus networks; ring structured networks, and tree-structured networks (figure 2). Some of the traditional distributed systems such as client/server paradigm use a star as the network topology. The problem with such a system is that when the central node fails, the entire system will be collapsed. In a multi-access bus network, a set of clients are connected via a shared communications line, called a bus. The bus link becomes the bottleneck and if it fails, all the nodes in the system cannot connect to each other. Another disadvantage is that performance degrades as additional computers are added or on heavy traffic. In a ring network each node connects to exactly two other nodes, forming a single continuous pathway for signals through each node. As new nodes are added, the diameter of the system grows as the number of nodes in the system, resulting in a longer message transmission delay. A node failure or cable break might isolate every node attached to the ring. In a tree-structured network (hierarchical network), the nodes are connected as a tree. Each node in the network having a specific fixed number, of nodes associated to it at the next lower level in the hierarchy. The scalability of the tree-structured network is better than that of the fully connected network, since new node can be added as the child node of the leaf nodes or the interior nodes. On the other hand, in such systems, only messages transmitted between a parent node and its child node go though one link, other messages transmitted between two nodes have to go through one or more intermediate nodes.

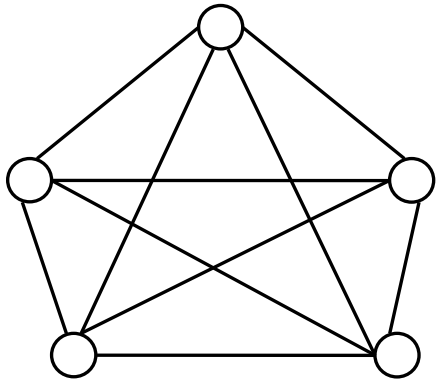
Fully Connected Network

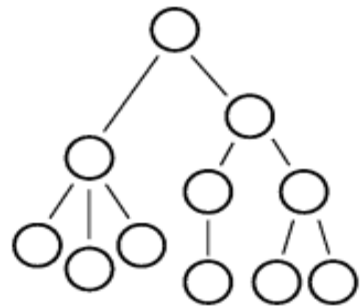
Tree Structured Network

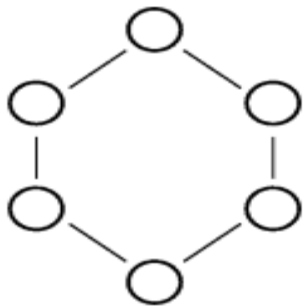
Ring Structured Network

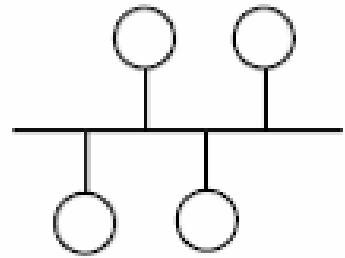
Multi-access Bus Network

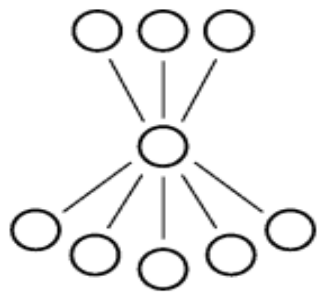
Star Structured Network

**Figure 2. Network Models**

## Characteristics of a Distributed System

A distributed system must possess the following characteristics to deliver utmost performance for the users:

- **Fault-Tolerant:** Distributed systems consist of a large number of hardware and software modules that are bound to fail in the long run. Such component failures can escort to service unavailability. Hence, the systems should be able to recover from component failures without performing erroneous actions. The goal of fault tolerance is to avoid failures in the system even in the presence of faults to provide uninterrupted service. A system is said to be fault tolerant if it can mask the presence of faults. The aim of any fault tolerant system is to increase its reliability or availability. The reliability of a system is defined as the probability that the system survives till that time. A reliable system prevents loss of information even in the event of component failures. Availability is the fraction of time for which a system is available for use. Usually fault tolerance is achieved by providing redundancy. Redundancy is defined as those parts of the system that are not needed for its correct functioning. It is of three types – hardware, software and time. Hardware redundancy is achieved by adding extra hardware components to system which take over the role of failed components in case some faults occur in them. Software redundancy includes extra instructions and code included for managing the extra hardware components, and using them correctly for uninterrupted service, in case of some component failure. In time redundancy the same instruction is executed many times. This is used to handle temporary faults in the system [VIKA04].

- **Scalable:** A distributed system can operate correctly even as some aspect of the system is scaled to a larger size. Scale has three components: the number of users and other entities that are part of the system, the distance between the farthest nodes in the system, and the number of organizations that exert administrative control over pieces of the system. The three elements of scale affect distributed systems in many ways. Among the affected components are naming, authentication for verifying someone's identity, authorization, communication, the use of remote resources, and the mechanisms by which users observe the system. Three techniques are employed to manage scale: *replication, distribution, and caching* [CLIF94].

Replication creates multiple copies of resources. Its use in naming, authentication, and file services reduces the load on individual servers and improves the reliability and availability of the services as a whole. The two important issues of replication are the placement of the replicas and the mechanisms by which they are kept consistent. The placement of replicas in a distributed system depends on the purpose for replicating the resource. If a service is being replicated to reduce the network delays when the service is accessed, the replicas are sprinkled across the system. If the majority of users are local, and if the service is being replicated to improve its availability or to spread the load across multiple servers, then replicas may be placed near one another. If a change is made to the object, the change should be noticeable to everyone in the system. For example, the system sends the updates to any replica, and that replica forwards the update to the others as they become available. If inconsistent updates are received by different replicas in different orders, timestamps (the date/time at which the update was generated) are used to differentiate the copies.

Distribution, another mechanism for managing scale in distributed systems, allows the information maintained by a distributed service to be extended across several servers. Distributing data across multiple servers reduces the size of the database that must be maintained by each server, dropping the time needed to search the database. Distribution also spreads the load across the servers reducing the number of requests that are handled by each. If requests can be distributed to servers in proportion to their power, the load on servers can be effectively managed. Network traffic can be reduced if data are assigned to servers close to the location from which they are most frequently used. In tree structured system, if cached copies are available from subordinate servers, the upper levels can be avoided.

Caching is another important technique for building scalable systems. Caching decreases the load on servers and the network. Cached data can be accessed faster than if a new request is made. The difference between replication and caching is that cached data is a short-term data. Instead of propagating updates on cached data, consistency is maintained by nullifying cached data when consistency can not be guaranteed. Caching is usually performed by the client, reducing frequent requests to network services. Caching can also occur on the servers executing those services.

Reading a file from the memory cached copy on the file server is faster than reading it from the client's local disk.

- **Predictable Performance:** Various performance metrics such as response time (elapsed time between the end of an inquiry or demand on a computer system and the beginning of a response), throughput (the rate at which a network sends or receives data), system utilization, network capacity etc. are employed to assess the performance. Predictable performance is the ability to provide desired responsiveness in a timely manner.

- **Openness:** The attribute 'openness' ensures that a subsystem is continually open to interaction with other systems. Web services are software systems designed to support interoperable machine-to-machine interaction over a network. These protocols allow distributed systems to be extended and scaled. An open system that scales has benefit over a completely closed and self-reliant system. A distributed system independent from heterogeneity of the underlying environment such as hardware and software platforms achieves the property of openness. Therefore, every service is equally accessible to every client (local or remote) in the system. The implementation, installation and debugging of new services should not be very complex in a system possessing openness characteristic.

- **Security:** Distributed systems should allow communication between programs/users/ resources on different computers by enforcing necessary security arrangements. The security features are mainly intended to provide confidentiality, integrity and availability. Confidentiality (privacy) is protection against disclosure to unauthorised person. Violation of confidentiality range from the discomforting to the catastrophic. Integrity provides protection against alteration and corruption. Availability keeps the resource accessible. Many incidents of hacking compromise the integrity of databases and other resources. "Denial of service" attacks are attacks against availability. Other important security concerns are access control and nonrepudiation. Maintaining access control facilitates the users to access only those resources and services to which they are entitled. It also ensures that users are not denied resources that they legitimately can expect to access. Nonrepudiation provides protection against denial by one of the entities involved in a communication. The security mechanisms put into practice should guarantee appropriate use of resources by different users in the system.

- **Transparency:** Distributed systems should be perceived by users and application developers as a whole rather than as a collection of cooperating components. The locations of the computer systems involved in the operations, concurrent operations, data replication, resource discovery from multiple sites, failures, system recovery etc. are hidden from users. Transparency hides the distributed nature of the system from its users and shows the user that the system is appearing and performing as a normal centralized system. The transparency can be employed in different ways in a distributed system (Figure 3) [KAZI00, PRAD02].

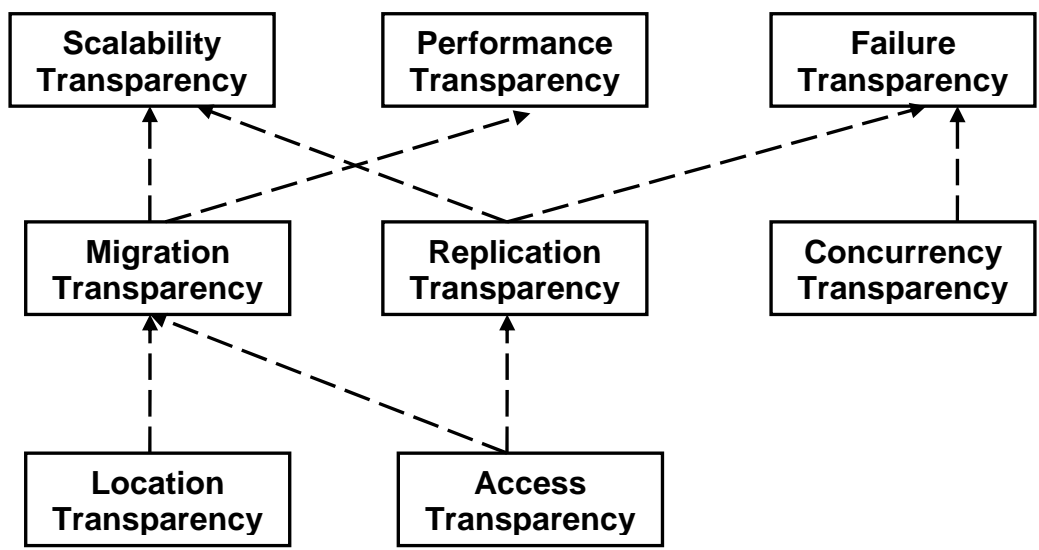

**Figure 3. Transparency in Distributed Systems**

- *Access transparency* facilitates the users of a distributed system to access local and remote resources using identical operations. (e.g. navigation in the web).
- *Location transparency* describes names used to identify network resources (e.g. IP address) independent of both the user's location and the resource location. In other words, location transparency facilitates a user to access resources from anywhere on the network without knowing where the resource is located. A file could be on the user's own PC, or thousands of miles away on other servers.

- *Concurrency transparency* enables several processes to operate concurrently using shared information objects without interference between them (e.g.: Automatic Teller Machine network). The users will not notice the existence of other users in the system (even if they access the same resources).
- *Replication transparency* enables the system to make additional copies of files and other resources for the purpose of performance and/or reliability, without the users noticing. If a resource is replicated among several locations, it should appear to the user as a single resource (e.g. Mirroring - Mirror sites are usually used to offer multiple sources of the same information as a way of providing reliable access to large downloads).
- *Failure transparency* enables the applications to complete their task despite failures occurring in certain components of the system. For example, if a server fails, but users are automatically redirected to another server and the user never notices the failure, the system is said to show high failure transparency. Failure transparency is one of the most difficult types of transparency to accomplish since it is hard to determine whether a server has actually failed, or whether it is simply responding very slowly. Moreover, it is generally unfeasible to achieve full failure transparency in a distributed system since networks are unreliable.
- *Migration transparency* facilitates the resources to move from one location to another without having their names changed. (e.g.: Web Pages). Users should not be aware of whether a resource or computing entity possesses the ability to move to a different physical or logical location.
- *Performance transparency* ensures the load variation should not lead to performance degradation. This could be achieved by automatic reconfiguration as response to changes of the load. (e.g.: load distribution)
- *Scalability transparency* allows the system to remain efficient even with a significant increase in the number of users and resources connected (e.g. World-Wide-Web, distributed database).

# The Eight fallacies of distributed computing

In 1994, Peter Deutsch of SUN drafted 7 assumptions [ARNO06, INGR04], architects and designers of distributed systems are expected to make, which prove wrong in the end. In 1997 James Gosling - father of the Java programming language, added another such fallacy – the eighth fallacy of distributed computing. The assumptions are now collectively known as the "The 8 fallacies of distributed computing". The fallacies of distributed Computing are a set of common but false assumptions made by programmers when first developing distributed applications. Many distributed systems which were developed based on these assumptions were needlessly complex caused by mistakes that required patching later on.

<u>The Fallacies of Distributed Computing</u>
1. The network is reliable.
2. Latency is zero.
3. Bandwidth is infinite.
4. The network is secure.
5. Topology doesn't change.
6. There is one administrator.
7. Transport cost is zero.
8. The network is homogeneous.

*Reliability:* The software which has been developed with the assumption that network is reliable; the network will lead to trouble when it starts dropping packets. Reliability can often be improved by increasing the autonomy of the nodes in a system. Replication can also improve the reliability of a system.

*Latency:* Latency is the time between initiating a request for data and the beginning of the actual data transfer. Latency can be comparatively good on a LAN; however it deteriorates quickly the user move to WAN scenarios or internet scenarios. Assuming latency is zero will definitely lead to scalability problems as the application grows geographically, or is moved to a different kind of network.

***Bandwidth:*** A measure of the capacity of a communications channel, i.e. how much data we can transfer during that time. The higher a channel's bandwidth, the more information it can carry. However, there are two forces at work to keep this assumption a fallacy. One is that while the bandwidth grows, so does the amount of information we try to squeeze through it. VoIP, videos, and IPTV are some of the newer applications that take up bandwidth. The other force at work to lower bandwidth is packet loss (along with frame size). Bandwidth limitations direct us to strive to limit the size of the information we send over the wire.

***Security:*** The network is never secure since the systems are facing various types of threats. Hence, the developer should perform threat modelling to evaluate the security risks. Following this, analyze which risk should be mitigated by what measures (a tradeoff between costs, risks and their probability) and take appropriate measures. Security is typically a multi-layered solution that is handled on the network, infrastructure, and application levels. The software architect should be conscious that security is very essential and the consequences it may have.

***Topology:*** Topology deals with the different configurations that can be adopted in building networks, such as a ring, bus, star or fully connected. For example any given node in the LAN will have one or more links to one or more other nodes in the network and the mapping of these links and nodes onto a graph results in a geometrical shape that determines the physical topology of the network. Likewise, the mapping of the flow of data between the nodes in the network determines the logical topology of the network. The physical and logical topologies might be identical in any particular network but they also may be different. When a new application is deployed in an organization, the network structure may also be altered. The operations team is likely to add and remove servers every once in a while and/or make other changes to the network. Finally, there are server and network faults which can cause routing changes. At the client's side the situation is even worse. There are laptops coming and going, wireless adhoc networks, new mobile devices etc. In a nutshell, topology in a distributed system is changing persistently [NELS01].

***Administrator:*** The sixth distributed computing fallacy is "there is one administrator". A simple situation is that with different administrators assigned according to expertise - databases, web servers, networks, different operating systems and the like for a company. The problem occurs when the company collaborates with external entities such as a business partner, or if the application is deployed for Internet consumption and hosted by some hosting service and the application consumes external services. In these situations, the other administrators are not even under company administrators control and they may have their own rules for administration. Hence the assumption of 'one administrator' is proven to be a myth.

Most of the time, the administrators are not part of the software development team. Therefore, the developers should provide them with tools to diagnose and find problems. A practical approach is to include tools for monitoring ongoing operations as well; for instance, to allow administrators recognize problems when they are small before they become a system failure. As a distributed system grows, its various components such as users, resources, nodes, networks, etc. will start to cross administrative domains. This means that the number of organizations or individuals that exert administrative control over the system will grow. In a system that scales poorly with regards to administrative growth, this can lead to problems of resource usage, reimbursement, security, etc.

*Transport cost:* Transport cost never becomes zero. The costs for setting and running the network are not free. There are costs for buying the routers, costs for securing the network, costs for leasing the bandwidth for internet connections, and costs for operating and maintaining the network running.

*Homogeneous network:* The eighth distributed computing fallacy is "network is homogeneous." Homogeneous network is a network derived of computers using similar configuration and protocols. Except a few very trivial ones, no network is homogeneous. Proprietary protocols are very harder to integrate. Hence, make use of standard technologies that are widely accepted such as XML (extended markup language) or Web Services as these technologies help alleviate the affects of the heterogeneity of the enterprise environment.

## Client/Server Computing

As networks of computing resources have become widespread, the notion of distributing interrelated processing amongst several resources has become popular. Over the years, numerous methods have evolved to facilitate this distribution. One of the popular distributed models is client/server computing [SILV98]. The client/server model is an extension of the modular programming model. Modular programming breaks down the design of a program into individual modules that can be programmed and tested independently. A modular program consists of a main module and one or more auxiliary modules. Like modular programming model, a client/server model consists of clients and servers. The clients and servers normally run on different computers interconnected by a computer network. The calling component becomes the client and the called component the server.

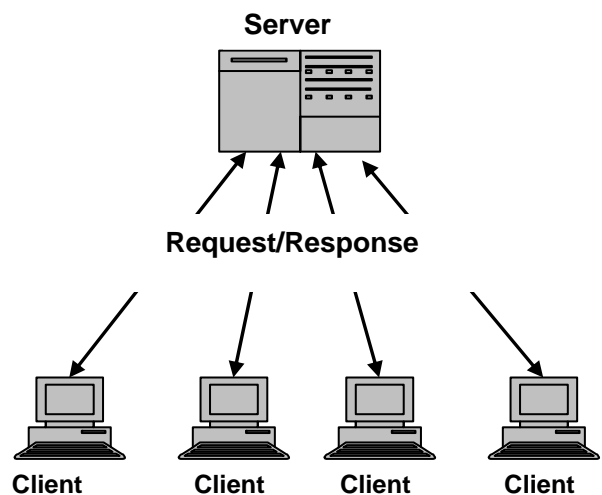

**Figure 4. Client/server communication**

A client application sends messages to a server via the network to request the server for performing a specific task. The client handles local resources such as input-output devices, local disks, and other peripherals. The server program listens for client requests that are transmitted via the network. Servers receive those requests and perform actions. Most of the data is processed on the server and only the results are returned to the client. This reduces the amount of network traffic between the server and the client machine. Thus network performance is improved further. The server controls the allocation of the information and also optimizes the resource consumption.

An important design consideration for large client/server systems is whether a client talks directly to the server, or whether an intermediary process is introduced between the client and the server. The former is a two-tier architecture (figure 4); the latter is a three-tier architecture. N-tier architecture is usually used for web applications to forward the requests further to other enterprise services. The two-tier architecture is easier to implement and is typically used in small environments. However, two-tier architecture is less scalable than a three-tier architecture.

In the three-tier architecture (figure 5), an intermediate process connects the clients and servers [CHAN05, JI96]. The intermediary can accumulate frequently used server data to guarantee enhanced performance and scalability. In database based 3-tier client/server architecture, normally there are three essential components: a client computer, an application server and a database server. The application server is the middle tier server which runs the business application. The data is retrieved from database server and it is forwarded to the client by the middle tier server. Middleware is a key to developing three-tier client/server application. Database-oriented middleware offers an Application Program Interface (API) access to a database. Java Database Connectivity (JDBC) is a well-known API, these classes can be inured to aid an applet or a servlet access any number of databases without considerate the inhabitant features of the database.

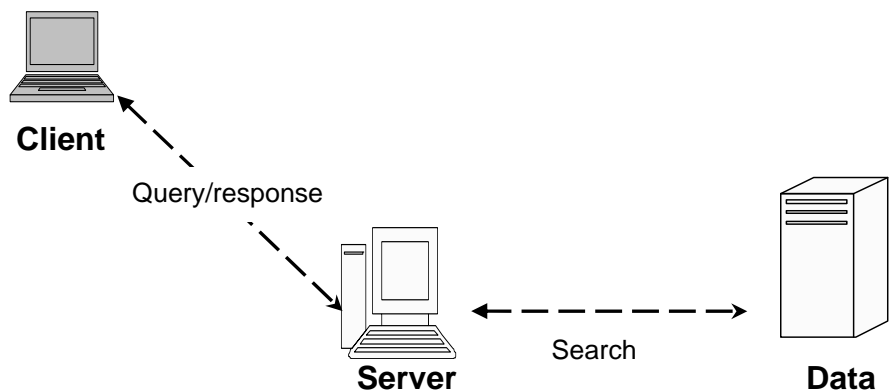

**Figure 5. 3-tier Client/server structure**

For security purposes servers must also address the problem of authentication. In a networked environment, an unauthorized client may attempt to access sensitive data stored on a server. Authentication of clients is provided by cryptographic techniques such as public key encryption or special authentication servers. Sometimes critical servers are replicated in order to achieve high availability and fault tolerance. If one replica fails then the other replicas hosted in different servers still remain available for the clients.

In the 3-tier architecture, it is easier to modify or replace any tier without affecting the other tiers. The separation of application and database functionality achieves better load balancing. Moreover, necessary security guidelines can be put into effect within the server tiers without hampering the clients.

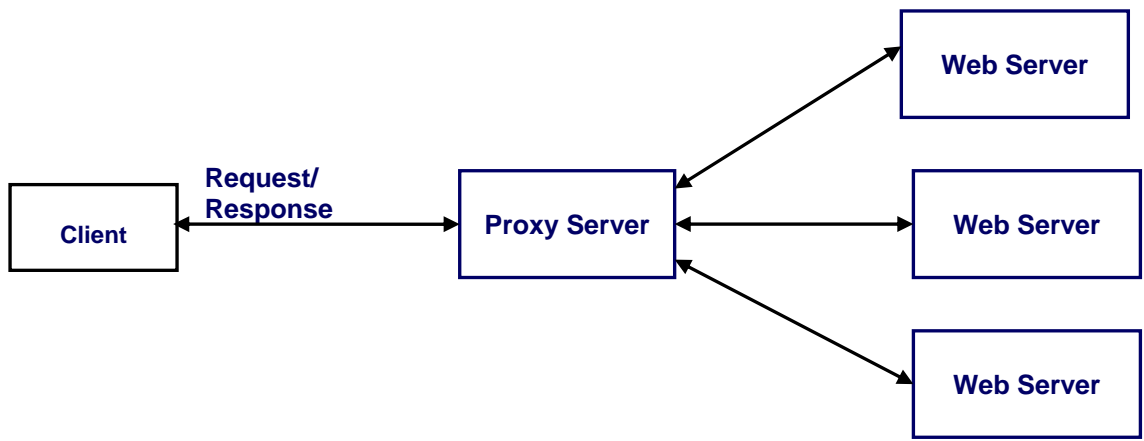

**Figure 6. Proxy Server Model**

A 3-tier client/server model known as 'proxy server model' (figure 6) is commonly used to improve retrieval performance of Internet. The intermediate process – proxy server, distributes client requests to several servers so that requests execute in parallel. A client connects to the proxy server, requesting some service, such as web page available in a web server. The proxy server assesses the request based on its filtering policy. For example, it may filter traffic by IP address or protocol. If the request is authenticated by the filter, the proxy presents the resource by connecting to the appropriate server and demanding the required service for the client. A proxy server may sometimes serve the request without contacting the specified web server. This is made possible by keeping the pages commonly visited by users in the cache of the proxy. By keeping local copies of frequently accessed file, the proxy can serve those files back to a requesting browser without going to the external site each time; this dramatically improves the performance seen by the end user. A proxy server with the ability to cache information is generally called a "proxy-cache server". A proxy is sometimes used to authenticate users by asking them to identify themselves, such as with a username and password. It is also easy to grant access to external resources only to authorised users, and to record each use of external resources in log files.  A proxy can also be used in a reverse direction to balance the load amongst a set of identical servers.

*Advantages of client/server model*

i. All resources are centralised, hence, server can manage resources that are common to all users. For example a central database would be used to evade problems caused by redundant and conflicting data.
ii. Improved security is offered as the number of entry points giving access to data is not so important.
iii. Server level administration is adequate as clients do not play a major role in the client/server model, they require less administration.
iv. A scalable network as it is possible to remove or add clients without affecting the operation of the network and without the need for major modification.

*Disadvantages of the client/server model*

i. Increased cost due to the technical complexity of the server.
ii. If a server fails, none of its clients will get service, unless the system is designed to be fault-tolerant.
iii. If the network fails, all servers become unreachable.
iv. If one client produces high network traffic, then all clients may suffer from long response times.

## World Wide Web– A massive distributed system

The Internet - a massive network of networks, connects millions of computers together worldwide, forming a network in which any computer can communicate with any other computer provided that they are both connected to the Internet. The World Wide Web (WWW), or simply Web, is a way of accessing information over the medium of the Internet. WWW consists of billions of web pages, spread across thousands and thousands of servers all over the world. It is an information-sharing model that is built on top of the Internet. The most well-known example of a distributed system is the collection of web servers. Hypertext is a document containing words that bond to other documents in the Web. These words are known as links and are selectable by the user. A single hypertext document can hold links to many documents.

The backbone of WWW are its files, called pages or Web pages, containing information and links to resources - both text and multimedia - throughout the Internet. Internet protocols are sets of rules that allow for inter-machine communication on the Internet. HTTP (HyperText Transfer Protocol) transmits hypertext over networks. This is the protocol of the Web. Simple Mail Transport Protocol or SMTP distributes e-mail messages and attached files to one or more electronic mailboxes. VoIP (Voice over Internet Protocol) allows delivery of voice communications over IP networks, for example, phone calls. A web server accepts HTTP requests from clients, and serving them HTTP responses along with optional data contents such as web pages.

The operation of the web relies primarily on hypertext as its means of information retrieval. Web pages can be created by user activity. Creating hypertext for the Web is accomplished by creating documents with a language called hypertext markup language, or HTML. With HTML, tags are placed within the text to achieve document formatting, visual features such as font size, italics and bold, and the creation of hypertext links. Servers implementing the HTTP protocol jointly provide the distributed database of hypertext and multimedia documents. The clients access the web through the browser software installed on their system. The URL (uniform resource locator) indicates the internet address of a file stored on a host computer, or server, connected to the internet. URLs are translated into numeric addresses using the domain name system (DNS). The DNS is a worldwide system of servers that stores location pointers to web sites. The numeric address, called the IP (Internet Protocol) address, is actually the "real" URL. Once the translation is made by the DNS, the browser can contact the web server and ask for a specific file located on its site. Web browsers use the URL to retrieve the file from the server. Then the file is downloaded to the user's computer, or client, and displayed on the monitor connected to the machine. Due to this correlation between clients and servers, the web is a client-server network. The web is used by millions of people everyday for different purposes including email, reading news, downloading music, online shopping or simply accessing information about anything. In fact, the web symbolizes a massive distributed system that materializes as a single resource to the user accessible at the click of a button. In order for the web to be accessible to anyone, some agreed-upon standards must be pursued in the formation and delivery of its content. An organization leading the efforts to standardize the web is the World Wide Web (W3C) Consortium.

**Web Information Retrieval**

Web information retrieval [AMI08, AMY04, AMY06, DIRK05, MONI03, VENK97] is the process of searching the world's largest and linked document collection – the World Wide Web, for information most relevant to a user's query. The various challenges of information retrieval on the web are: (i) data is distributed - data spans over many computers, of a variety of platforms, (ii) data is volatile - computers and files are added and removed frequently and unpredictably, (iii) volume of data is very huge - growth continues exponentially, (iv) data quality is inconsistent - data may be false, error-ridden, invalid, outdated, ambiguous and multiplicity of sources introduces inconsistency and (v) heterogeneous data - multiple media types and media formats and multiple languages and alphabets. As a result, it would be physically unfeasible for an individual to sift through and examine all these pages to find the required information. Usually, in order to search for information on the internet a software tool called Search Engine is used. When a user enters a query into a search engine from their browser software, their input is processed and used to search the database for occurrences of particular keywords. A variety of search engines such as Google, Yahoo! Search, are available to make the web retrieval process very faster. Two main architectures used for web searching are *centralised and distributed search.*

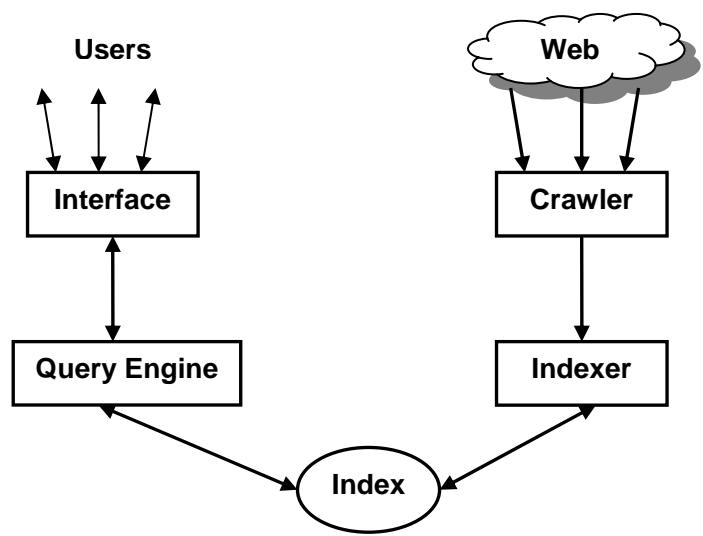

**Figure 7. Search Engine: Centralised Architecture**

*Centralised Architecture*: The aim of centralised approach is to index sizeable portion of Web, independently of topic and domain. The centralised architecture based search engine has main three parts: *a crawler, an indexer, and query handler*. The crawler (spider or robot) retrieves web pages, compress and store into a *page repository*. This process is called crawling (sometimes known as robot spidering, gathering or harvesting). Some of the most well known crawlers include Googlebot (from Google) MSNBot (from MSN) and Slurp (from Yahoo!). Crawlers are directed by a crawler control module that gives the URLs to visit next. The indexer processes the web pages collected by the crawler and builds an index, which is the main data structure used by the search engine and represents the crawled web pages. The inverted index contains for each word a sorted list of couples such as docID and position in the document. The *query engine* processes the user queries and returns matching results using the index. The results are returned to the user in an order determined by a ranking algorithm. Each search engine may have a different ranking algorithm, which parses the pages in the engine's database to determine relevant responses to search queries. Some search engines keep a local cache copy of many popular pages indexed in their database, to allow for faster access and in case the destination server is temporarily inaccessible.

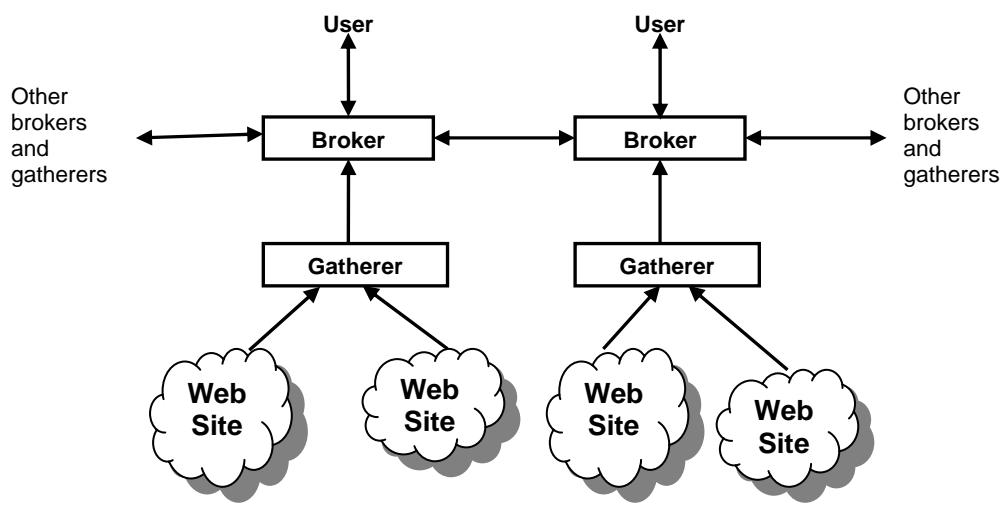

**Figure 8. Search Engine: Distributed Architecture**

*Distributed architecture:* Searching is a coordinated effort of many information gatherers and brokers. Gatherer extracts information (called summaries) from the documents stored on one or more web servers. It can handle documents in many formats: HTML, PDF, Postscript, etc. Broker obtains summaries from gatherers, stores them locally, and indexes the data. It can search the data; fetch data from other brokers and makes data available for user queries and to other brokers. The advantages of distributed architecture are the gatherer running on a server reduces the external traffic on that server and evading of gatherer sending information to multiple brokers reduces work repetition.

**Advanced Distributed System Models**

The distributed systems are cost-effective as compared to central systems. The introduction of redundancy amplifies the availability even some parts of the system stop working. Quite a few applications can be run concurrently in a distributed system. Adding new components does not influence the performance of distributed systems as the systems are scalable. A large number of computers take part in performing a distributed computing task. Thus, distributed systems provide shorter response time and superior throughput than centralised systems. Another merit is that distributed systems are very reliable. The distributed systems have the benefit of being highly available. Because of these multiple benefits, a range of distributed systems and applications have been developed recently and are being used extensively in the real world. Well-known distributed computing systems are clusters, grids, peer-to-peer networks (P2P), distributed storage systems [ANDR02, BUYY99, IAN99, POUR05] and so on. Moreover, mobile computing based distributed systems are also emerging.

The concept behind clustering, in its simplest form, is that many smaller computers can be grouped in a way to make a computing structure that could provide all of the processing power needed, for much less capital. A grid is a type of distributed system that allows coordinated sharing and aggregation of distributed, heterogeneous resources based on users' requirements. Grids are normally used to support applications emerging in the areas of e-Science and e-Business. Usually grid computing involves geographically distributed communities of people who engage in collaborative activities to solve large scale problems. This requires sharing of various resources such as computers, data, applications and scientific apparatuses. P2P networks are decentralized distributed systems, which enable applications such as file-sharing, instant messaging, internet telephony, content distribution, high performance computing etc... The most renowned function of peer-to-peer networking is found in file-exchanging communities. P2P technology is also used for applications such as the harnessing of the dormant processing power in desktop PCs and remote storage sharing. Distributed storage systems provide users with a unified view of data stored on different file systems and computers which may be on the same or different networks.

The advanced distributed system models are discussed in detail in the subsequent chapters of this book. Chapters 2, 3, 4 and 6 are dedicated to P2P networking. An overview of cluster and grid computing technologies are presented in Chapter 7. Other advanced topics in distributed computing area are discussed in chapter 8.

## Conclusions

In this chapter an overview of distributed systems are presented. The architecture, various characteristics, and the myths on distributed computing are discussed. Further to that, an introduction to client/server systems and World Wide Web are given. The future of distributed computing is still quite uncertain since it is one of many new types of computing. The technology has truly shown its worth as a useful tool for various complex applications.